\documentstyle [12pt] {article}

\catcode`@=12
\begin{document}
\thispagestyle{empty}
\begin{flushright}
{UCRHEP-T128\\July 1994\\}
\end{flushright}
\vspace{0.5in}
\begin{center}
{\Large \bf Scalar Mass Bounds in Two Supersymmetric\\
            Extended Electroweak Gauge Models\\}
\vspace{1.5in}
{\bf T. V. Duong and Ernest Ma\\}
\vspace{0.3in}
{\sl Department of Physics\\}
{\sl University of California\\}
{\sl Riverside, California 92521\\}
\vspace{1.5in}
\end{center}
\begin{abstract}
In two recently proposed supersymmetric extended electroweak gauge models,
the reduced Higgs sector at the 100-GeV energy scale consists of only two
doublets, but they have quartic scalar couplings different from those of
the minimal supersymmetric standard model.  In the $\rm SU(2) \times
SU(2) \times U(1)$ model, there is an absolute upper bound of about
145 GeV on the mass of the lightest neutral scalar boson.
In the $\rm SU(3) \times U(1)$ model, there is only a
parameter-dependent upper bound which formally goes to infinity in a
particular limit.
\end{abstract}

\newpage
\baselineskip 24pt

It has been demonstrated recently in two explicit examples\cite{1,2} that
in a supersymmetric extension of the standard $\rm SU(2)_L \times U(1)_Y$
electroweak gauge model, even though there may be only two scalar doublets at
the electroweak energy scale, they are not necessarily those of the
minimal supersymmetric standard model (MSSM).  The reason is that the
$\rm SU(2) \times U(1)$ gauge symmetry may be a remnant\cite{3} of
a larger symmetry which is broken at a higher mass scale together with
the supersymmetry.  The structure of the reduced Higgs potential is then
determined by the scalar particle content of the larger theory and there
are in general additional couplings appearing in the superpotential which
contribute to the quartic couplings of the two Higgs doublets at the
electroweak energy scale.

Consider the Higgs potential $V$ for two scalar doublets $\Phi_{1,2} =
(\phi_{1,2}^+, \phi_{1,2}^0)$:
\begin{eqnarray}
V &=& \mu_1^2 \Phi_1^\dagger \Phi_1 + \mu_2^2 \Phi_2^\dagger \Phi_2 +
\mu_{12}^2 (\Phi_1^\dagger \Phi_2 + \Phi_2^\dagger \Phi_1) \nonumber \\
&+& {1 \over 2} \lambda_1 (\Phi_1^\dagger \Phi_1)^2 + {1 \over 2} \lambda_2
(\Phi_2^\dagger \Phi_2)^2 + \lambda_3 (\Phi_1^\dagger \Phi_1) (\Phi_2^\dagger
\Phi_2) \nonumber \\ &+& \lambda_4 (\Phi_1^\dagger \Phi_2) (\Phi_2^\dagger
\Phi_1) + {1 \over 2} \lambda_5 (\Phi_1^\dagger \Phi_2)^2 + {1 \over 2}
\lambda_5^* (\Phi_2^\dagger \Phi_1)^2.
\end{eqnarray}
In the MSSM, there are the well-known constraints
\begin{equation}
\lambda_1 = \lambda_2 = {1 \over 4} (g_1^2 + g_2^2), ~~ \lambda_3 =
-{1 \over 4} g_1^2 + {1 \over 4} g_2^2, ~~ \lambda_4 = -{1 \over 2} g_2^2,
{}~~ \lambda_5 = 0,
\end{equation}
where $g_1$ and $g_2$ are the U(1) and SU(2) gauge couplings of the
standard model respectively.  Note that only the gauge couplings contribute
to the $\lambda$'s.  This is because that with two $\rm SU(2) \times
U(1)$ Higgs superfields, there is no cubic invariant in the superpotential
and thus no additional coupling.

In the supersymmetric $\rm SU(2)_L \times SU(2)_R \times U(1)$ model based
on E$_6$ particle content,\cite{4} it has been shown\cite{1} that instead
of Eq. (2), the quartic scalar couplings of the two Higgs doublets are
now given by
\begin{eqnarray}
\lambda_1 &=& {1 \over 4} (g_1^2 + g_2^2) + {{2 f^2 g_1^2} \over g_2^2} -
{{4 f^4} \over g_2^2} \left( 1 - {g_1^2 \over g_2^2} \right), \\
\lambda_2 &=& {1 \over 4} (g_1^2 + g_2^2) + 2 f^2 \left( 1 - {g_1^2 \over
g_2^2} \right) - {{4 f^4} \over g_2^2} \left( 1 - {g_1^2 \over g_2^2}
\right), \\ \lambda_3 &=& -{1 \over 4} g_1^2 + {1 \over 4} g_2^2 + f^2 -
{{4 f^4} \over g_2^2} \left( 1 - {g_1^2 \over g_2^2} \right), \\
\lambda_4 &=& -{1 \over 2} g_2^2 + f^2, ~~~~~~ \lambda_5 ~=~ 0,
\end{eqnarray}
where $f$ is a coupling in the superpotential of the larger theory.\cite{1}
The general requirement that $V$ be bounded from below puts an upper limit
on $f$: namely
\begin{equation}
f_{max}^2 = {1 \over 4} (g_1^2 + g_2^2) \left( 1 - {g_1^2 \over g_2^2}
\right)^{-1},
\end{equation}
which corresponds to having soft supersymmetry breaking terms in the Higgs
sector which also preserve the left-right symmetry.  If $f = f_{max}$,
the lighter of the two neutral Higgs bosons is constrained by\cite{1}
\begin{equation}
m_h^2 < 2 M_W^2 \left[ 1 - {{2 x_W^2} \over {(1-x_W) (1-2x_W)}} \right]
\sin^4 \beta + \epsilon,
\end{equation}
where
\begin{equation}
\epsilon = {{3g_2^2m_t^4} \over {8\pi^2M_W^2}} \ln \left( 1 + {\tilde m^2
\over m_t^2} \right)
\end{equation}
is an extra term coming from radiative corrections\cite{5} and $x_W \equiv
\sin^2 \theta_W$, $\tan \beta \equiv \langle \phi_2^0 \rangle / \langle
\phi_1^0 \rangle$.  Using $m_t = 174$ GeV\cite{6} and assuming $\tilde m =
1$ TeV, this gives an upper bound of about 133 GeV on $m_h$ instead of about
128 GeV if the MSSM were assumed.  However, the choice of $f = f_{max}$
does not maximize the upper bound on $m_h$, as shown below.

For a Higgs potential with $\lambda_5 = 0$, the upper bound on the lighter
of the two neutral scalar Higgs bosons is given in general by
\begin{equation}
(m_h^2)_{max} = 2 (v_1^2 + v_2^2) [\lambda_1 \cos^4 \beta + \lambda_2
\sin^4 \beta + 2 (\lambda_3 + \lambda_4) \sin^2 \beta \cos^2 \beta ]
+ \epsilon,
\end{equation}
where $v_{1,2} = \langle \phi_{1,2}^0 \rangle$.  Using Eqs. (3) to (6),
we then obtain
\begin{eqnarray}
(m_h^2)_{max} = {1 \over 2} (v_1^2 + v_2^2) [ (g_1^2 + g_2^2)
\cos^2 2 \beta + 8 f^2 ( 1 - \cos^4 \beta + (g_1^2 / g_2^2)
\cos 2 \beta ) \nonumber \\ - 16 (f^4 / g_2^2) ( 1 - g_1^2 /
g_2^2) ] + \epsilon ,~~~~~~~~
\end{eqnarray}
which is maximized at
\begin{equation}
f_0^2 = {1 \over 4} g_2^2 \left( 1 - \cos^4 \beta + {g_1^2 \over g_2^2}
\cos 2 \beta \right) \left( 1 - {g_1^2 \over g_2^2} \right)^{-1}.
\end{equation}
Since $f_0$ never exceeds $f_{max}$, the former should be used to
maximize $(m_h^2)_{max}$ and we obtain
\begin{equation}
(m_h^2)_{max} < 2(v_1^2 + v_2^2) \left[ {1 \over 4} (g_1^2 + g_2^2)
\cos^2 2 \beta + {{g_2^2 (1-\cos^4 \beta + (g_1^2/g_2^2) \cos 2 \beta)^2}
\over {4 (1-g_1^2/g_2^2)}} \right] + \epsilon.
\end{equation}
We plot this upper bound (solid line) as a function of $\cos^2 \beta$
in Fig. 1.  The absolute upper bound is reached at $\cos^2 \beta = 0$
where it is equal to $(2 M_W^2 + \epsilon)^{1/2}$ which is about 145 GeV.
For comparison, we show the corresponding upper
bound (dashed line) for $f = f_{max}$.  We show also the upper bound
(dotted line) of the MSSM, namely $(M_Z^2 \cos^2 2 \beta +\epsilon)^{1/2}$,
corresponding to $f = 0$.

In the supersymmetric extension of the new $\rm SU(3) \times U(1)$
model\cite{7} the reduced Higgs potential of two scalar doublets at the
electroweak energy scale is obtained\cite{2} with the following quartic
scalar couplings:
\begin{eqnarray}
\lambda_1 &=& {1 \over 4} (g_1^2 + g_2^2) + \left( 1 + {{3 g_1^2} \over
g_2^2} \right) {x \over {1-y}} - {3 \over g_2^2} \left( 1 - {{3 g_1^2}
\over g_2^2} \right) {{(1+yr)x^2} \over {(1-y)^2}}, \\ \lambda_2 &=&
{1 \over 4} (g_1^2 + g_2^2) + \left( 1 - {{3 g_1^2} \over g_2^2} \right)
{x \over {1-y}} - {3 \over g_2^2} \left( 1 - {{3 g_1^2} \over g_2^2}
\right) {{(1+yr)x^2} \over {(1-y)^2}}, \\ \lambda_3 &=& -{1 \over 4} g_1^2
+ {1 \over 4} g_2^2 + {x \over {1-y}} - {3 \over g_2^2} \left( 1 -
{{3 g_1^2} \over g_2^2} \right) {{(1+yr)x^2} \over {(1-y)^2}}, \\ \lambda_4
&=& -{1 \over 2} g_2^2 + x, ~~~~~ \lambda_5 = 0,
\end{eqnarray}
where $x \equiv f^2$, $y \equiv (u'/u)^2$, and $r \equiv (M/M')^2$.
Here, $f$ is the extra coupling appearing in the superpotential of the
larger theory, $u$ and $u'$ are the vacuum expectation values of the two
$\rm SU(3) \times U(1)$ scalar triplets which break the gauge symmetry
down to the standard $\rm SU(2) \times U(1)$, $M'$ is the arbitrary mass
of the corresponding heavy pseudoscalar boson, and
\begin{equation}
M^2 = {2 \over 3} g_2^2 \left( 1 - {{3 g_1^2} \over g_2^2} \right)^{-1}
(u^2 + u'^2).
\end{equation}
Details have been given elsewhere.\cite{2}  Using Eq. (10) we then find
\begin{eqnarray}
(m_h^2)_{max} = {1 \over 2} (v_1^2 + v_2^2) [ (g_1^2 + g_2^2) \cos^2 2 \beta
+ 4 x (1-y)^{-1} \{ 1 + 3 (g_1^2/g_2^2) \cos 2 \beta \nonumber \\
+{1 \over 2} \sin^2 2 \beta (1-y) \} - 12 x^2 g_2^{-2} ( 1 - 3 g_1^2/g_2^2 )
(1+yr) (1-y)^{-2} ] + \epsilon.
\end{eqnarray}
In the above, $r$ appears only in one term and since that term is
necessarily negative, $(m_h^2)_{max}$ is always maximized with the
smallest possible value of $r$.  Hence we need only consider $r = 0$
from now on.

The requirement that $V$ be bounded from below implies that $\lambda_1
\geq 0$, $\lambda_2 \geq 0$, and either $\lambda_3 + \lambda_4 \geq 0$
or if $\lambda_3 + \lambda_4 < 0$, then $|\lambda_3 + \lambda_4| \leq
\sqrt {\lambda_1 \lambda_2}$.  We show in Fig. 2 the resulting upper
bound on $x$ as a function of $y$.  There are four separate regions:
\begin{eqnarray}
0 \leq y \leq 1, &{\rm then}& x \leq x_2 = {1 \over 6} g_2^2 (1-y) \left[
1 + {2 \over {(1 - 3 g_1^2/g_2^2)^{1/2}}} \right], \\
1 \leq y \leq 3, &{\rm then}& x = 0, \\ 3 \leq y \leq 4, &{\rm then}&
x \leq x_3, \\ y \geq 4, &{\rm then}& x \leq x_1 = {1 \over 6} g_2^2 (y-1).
\end{eqnarray}
In the above, $x_1$ and $x_2$ correspond to $\lambda_1 = 0$ and $\lambda_2
= 0$ respectively, whereas $x_3$ is given by the smaller of the two roots
of the quadratic equation $[\lambda_1 \lambda_2 - (\lambda_3 + \lambda_4)^2]/x
= 0$, namely
\begin{equation}
{3 \over g_2^2} \left( 1 - {{3g_1^2} \over g_2^2} \right) x^2 - \left[
1 - {{3g_1^2} \over g_2^2} + {1 \over 2} (y-2)^2 \right] x + {1 \over 4}
(g_1^2 + g_2^2) (y-1) (y-3) = 0.
\end{equation}
Previously\cite{2} we had considered only the special case $y = 0$
and further assumed that $(\lambda_3 + \lambda_4)^2 \leq \lambda_1 \lambda_2$
so that the pseudoscalar mass $m_A$ is always allowed to be zero.  If
$\lambda_3 + \lambda_4 \geq \sqrt {\lambda_1 \lambda_2}$, then $m_A$ has
a lower bound given by
\begin{equation}
m_A^2 \geq {{4 [ (\lambda_3 + \lambda_4)^2 - \lambda_1 \lambda_2 ] v_1^2
v_2^2} \over {(m_h^2)_{max}}}
\end{equation}
which is in principle not a problem.

For a given value of $y$, we now vary $x$ to maximize the value of
$(m_h^2)_{max}$.  Let $z \equiv \cos^2 \beta$, then
\begin{equation}
x_0 = {1 \over 6} g_2^2 (1-y) \left( 1 - {{3 g_1^2} \over g_2^2} \right)^{-1}
\left[ 1 + {{3 g_1^2} \over g_2^2} (2z-1) + 2z(1-z)(1-y) \right]
\end{equation}
would be the value of $x$ at which $(m_h^2)_{max}$ is maximized for
$0 \leq y < 1$, unless $x_0 > x_2$ in which case $x_2$ should be
used instead.  For a given value of $z$, it can be shown that $(m_h)_{max}$
is maximized at $y=0$ with $x_0$ for $z < 0.176$ and $x_2$ for $z > 0.176$.
We plot in Fig. 3 this upper bound as a function of $z$, as well as lines
for some other values of $y$.  For any value in the range $0 \leq y < 1$,
$(m_h^2)_{max}$ is maximized at $z = 1$ where it is equal to
\begin{equation}
2 g_1^2 (v_1^2 + v_2^2) \left[ 1 + {2 \over {(1 - 3g_1^2/g_2^2)^{1/2}}}
\right] + \epsilon.
\end{equation}
Hence $m_h$ has an absolute upper bound of about 258 GeV in this interval.
However, it should be noted that near $y = 1$, our
approximation for the reduced Higgs potential breaks down because one of
the mass eigenvalues of the heavy neutral scalar mass matrix becomes too
small.  On the other hand, this problem does not affect the upper bound
on $(m_h)_{max}$ for any given value of $z$ because it always occurs at
$y = 0$ as shown in Fig. 3.

In the interval $1 \leq y \leq 3$, since $x = 0$ is required, this model
is identical to the MSSM in its reduced scalar sector.  Hence $m_h$ is
bounded by the dotted line of Fig. 1.

In the interval $3 \leq y \leq 4$, we need to consider the lines $x_0 = 0$
and $x_0 = x_3$ in the $(y,z)$ plane.  For a given value of $y$,
$(m_h)_{max}$ is maximized with $x=0$ for $z < z_1$ and $z > z_2$,
where $z_{1,2}$ are the roots of the quadratic equation
\begin{equation}
z^2 - z \left( 1 - {{3 g_1^2/g_2^2} \over {y-1}} \right) + {{1-3g_1^2/
g_2^2} \over {2(y-1)}} = 0,
\end{equation}
with $x=x_0$ for $z_1 < z < z_3$ and $z_4 < z < z_2$, and with $x=x_3$
for $z_3 < z < z_4$, where $z_{3,4}$ are the roots of the quadratic
equation
\begin{equation}
z^2 - z \left( 1 - {{3 g_1^2/g_2^2} \over {y-1}} \right) + {{1-3g_1^2/
g_2^2} \over {2(y-1)}} \left( 1 + {{6 x_3/g_2^2} \over {y-1}} \right)
= 0.
\end{equation}
We plot in Fig. 4 this upper bound as a function of $z$ for various values
of $y$.

For $y \geq 4$, we need to consider the lines $x_0=0$ and $x_0 = x_1$.
For a given value of $y$, $(m_h)_{max}$ is maximized with $x=0$ again
for $z < z_1$ and $z > z_2$, with $x=x_0$ now for $z_1 < z < z_5$ and
$z_6 < z < z_2$, and with $x=x_1$ for $z_5 < z < z_6$, where $z_{5,6}$
are the roots of the quadratic equation
\begin{equation}
z^2 - z \left( 1 - {{3g_1^2/g_2^2} \over {y-1}} \right) + {{1-3g_1^2/
g_2^2} \over {y-1}} = 0.
\end{equation}
We plot in Fig. 5 this upper bound on $(m_h)_{max}$ as a function of $z$
for various values of $y$.  We find that for $y - 4 > 6g_1^2/g_2^2$,
$(m_h^2)_{max}$ at $x = x_1$ has a local maximum at
\begin{equation}
z = {1 \over 2} \left[ 1 - {{3 g_1^2/g_2^2} \over {y - 4 -3 g_1^2/g_2^2}}
\right],
\end{equation}
where it is equal to
\begin{equation}
{1 \over 6} g_2^2 (v_1^2 + v_2^2) {{(y - 4)^2} \over
{y-4-3g_1^2/g_2^2}} + \epsilon.
\end{equation}
It is clear that this expression is unbounded as $y$ increases and it
exceeds $M_Z^2 + \epsilon$ at $y \simeq 6.5$, and the upper bound (27) at
$y \simeq 30.5$. However, as Eq. (23) shows, this requires the Yukawa
coupling $f$ to be increasingly large.

In summary, we plot in Fig. 6 the upper bound on $m_h$ as a function of $y$.
In the interval $0 \leq y < 1$, it is about 258 GeV, then it has the MSSM
value of 128 GeV from $y=1$ to $y \simeq 6.5$, after which it grows
according to Eq. (32).  Details are shown in Figs. 2 to 5.

The experimental search for Higgs bosons is an active ongoing effort.\cite{8}
It should be understood that even if supersymmetry exists and there are only
two Higgs doublets at the electroweak energy scale, they are not necessarily
those of the MSSM.  In particular, the upper bound on $m_h$ may exceed that
predicted by the MSSM, as shown here in two specific examples.

\newpage
\begin{center}{ACKNOWLEDGEMENT}
\end{center}

This work was supported in part by the U. S. Department of Energy under
Grant No. DE-FG03-94ER40837.

\bibliographystyle{unsrt}

\newpage
\begin{center} {\large \bf Figure Captions}
\end{center}

\begin{description}
\item[Fig.~1] Upper bound (solid line) on $m_h$ as a function of $\cos^2
\beta$ in the supersymmetric $\rm SU(2)_L \times SU(2)_R \times U(1)$
model based on E$_6$ particle content.  The dashed and dotted lines refer
to the special cases where $f = f_{max}$ and $f = 0$ respectively.
\item[Fig.~2] Upper bound on $x \equiv f^2$ as a function of $y \equiv
(u'/u)^2$ in the new supersymmetric $\rm SU(3) \times U(1)$ model.
\item[Fig.~3] Upper bound on $m_h$ as a function of $z \equiv \cos^2 \beta$
in the interval $0 \leq y \leq 1$ for various values of $y$.
\item[Fig.~4] Upper bound on $m_h$ as a function of $z$ in the interval
$3 \leq y \leq 4$ for various values of $y$.
\item[Fig.~5] Upper bound on $m_h$ as a function of $z$ for various values
of $y > 4$.
\item[Fig.~6] Upper bound on $m_h$ as a function of $y$.
\end{description}
\end{document}